\documentclass[aps,twocolumn,showpacs,preprintnumbers,prb,amsmath,amssymb,amsfonts,superscriptaddress,floatfix]{revtex4-1}

\usepackage[latin1]{inputenc}
\usepackage{graphics}
\usepackage{color}
\usepackage{amssymb}
\usepackage{amsmath}
\usepackage{overpic}
\usepackage{epstopdf}
\usepackage{bm}
\usepackage{graphicx}
\usepackage{subfigure}

\begin{document}

\affiliation{Key Laboratory of Automobile Materials of MOE and College of Materials Science and Engineering, Jilin University, Changchun 130012, China}
\affiliation{Department of Chemistry and Physics, Arkansas State University, AR 72467, USA}
\affiliation{National Laboratory of Solid State Microstructures, School of Electronic Science and Engineering and Collaborative Innovation Center of Advanced Microstructures, Nanjing University, Nanjing 210023, China}
\affiliation{State Key Laboratory of Superhard Materials, Jilin University, Changchun 130012, China}

\title{InSe: a two-dimensional material with strong interlayer coupling}

\author{Yuanhui Sun}
\affiliation{Key Laboratory of Automobile Materials of MOE and College of Materials Science and Engineering, Jilin University, Changchun 130012, China}
\author{Shulin Luo}
\affiliation{Key Laboratory of Automobile Materials of MOE and College of Materials Science and Engineering, Jilin University, Changchun 130012, China}
\author{Xin-Gang Zhao}
\affiliation{Key Laboratory of Automobile Materials of MOE and College of Materials Science and Engineering, Jilin University, Changchun 130012, China}
\author{Koushik Biswas}
\affiliation{Department of Chemistry and Physics, Arkansas State University, AR 72467, USA}
\author{Song-Lin Li}
\email{sli@nju.edu.cn}
\affiliation{National Laboratory of Solid State Microstructures, School of Electronic Science and Engineering and Collaborative Innovation Center of Advanced Microstructures, Nanjing University, Nanjing 210023, China}
\author{Lijun Zhang}
\email{lijun$\_$zhang@jlu.edu.cn}
\affiliation{Key Laboratory of Automobile Materials of MOE and College of Materials Science and Engineering, Jilin University, Changchun 130012, China}
\affiliation{State Key Laboratory of Superhard Materials, Jilin University, Changchun 130012, China}

\date{\today}

\begin{abstract}
Atomically thin, two-dimensional (2D) indium selenide (InSe) has attracted 
considerable attention due to large tunability in
the band gap (from 1.4 to 2.6 eV) and high carrier mobility. 
The intriguingly high dependence of band gap on layer
thickness may lead to novel device applications, although its origin remains poorly understood, 
and generally attributed to quantum confinement effect. 
In this work, we demonstrate via first-principles calculations that strong interlayer 
coupling may be mainly responsible for this phenomenon, especially in the fewer-layer region, 
and it could also be an essential factor influencing other material properties of $\beta$-InSe and $\gamma$-InSe. 
Existence of strong interlayer coupling manifests itself in three aspects: 
(i) indirect-to-direct band gap transitions with increasing layer thickness; 
(ii) fan-like frequency diagrams of the shear and breathing modes of few-layer flakes; 
(iii) strong layer-dependent carrier mobilities. 
Our results indicate that multiple-layer InSe may be deserving of 
attention from FET-based technologies and also an ideal system to study interlayercoupling, 
possibly inherent in other 2D materials.


\end{abstract}

\maketitle

\section{\textbf{INTRODUCTION}}

Isolation of graphene \cite{Novoselov666,geim_rise_2007,meyer_structure_2007} has triggered broad 
interest in the investigation of two-dimensional (2D) layered materials, 
such as hexagonal boron nitride (h-BN), \cite{satta_ab_2001,ishii_growth_1983,arya_preparation_1988} transition metal 
dichalcogenides (TMDCs)\cite{radisavljevicb._single-layer_2011,novoselov_two-dimensional_2005,yoon_how_2011,cai_polarity-reversed_2014} 
and black phosphorus (BP).\cite{li_quantum_2016,qiao_high-mobility_2014,hu_interlayer_2016} 
Recently, few-layer InSe as a post-transition metal chalcogenide has been synthesized via 
physical\cite{feng_back_2014,tamalampudi_high_2014,beardsley_nanomechanical_2016,brotons-gisbert_nanotexturing_2016,mudd_tuning_2013,bandurin_high_2017} and chemical methods\cite{lei_evolution_2014}, exhibits promising characteristics for 
optoelectronic applications.\cite{tamalampudi_high_2014,lei_evolution_2014,feng_ultrahigh_2015,mudd_high_2015,balakrishnan_room_2014,lei_optoelectronic_2015,camassel_excitonic_1978,huo_optoelectronics_2017} High tunability in the band gap with 
varying layer thickness has been confirmed by experimental and theoretical investigations.\cite{brotons-gisbert_nanotexturing_2016,mudd_tuning_2013,bandurin_high_2017,lei_evolution_2014,zolyomi_electrons_2014,errandonea_crystal_2005,gomes_da_costa_first-principles_1993,sanchez-royo_electronic_2014,sun_ab_2016,debbichi_two-dimensional_2015} In addition, carrier mobilities reaching or exceeding $10^3 cm^2V^{-1}s^{-1}$ at room temperature have been reported,\cite{feng_back_2014,bandurin_high_2017,segura_electron_1984,sucharitakul_intrinsic_2015} which are the highest among TMDCs\cite{cui_multi-terminal_2015,fallahazad_shubnikov-haas_2016} and are superior to black phosphorus.\cite{li_quantum_2016,qiao_high-mobility_2014}

Previous experimental and theoretical studies\cite{mudd_tuning_2013,bandurin_high_2017,lei_evolution_2014,gomes_da_costa_first-principles_1993,yang_spectroscopy_2005,yang_transient_2006,e_kress-rogers_and_g_f_hopper_and_r_j_nicholas_and_w_hayes_and_j_c_portal_and_a_chevy_electric_1983} 
on 2D InSe have already illustrated the evolution of electronic structure with thickness, 
which is generally attributed to quantum confinement effect,\cite{mudd_tuning_2013,yang_spectroscopy_2005,yang_transient_2006} 
a common feature observed in nanomaterials. In this paper, we calculate the electronic structure of $\beta$-InSe and $\gamma$-InSe 
by employing ab initio density functional (DFT) calculations with long-range dispersion interaction in order to analyse broad variations in the band gap with changing layer thickness. We find that interlayer coupling in layered InSe is not as weak as generally believed to be the case in conventional van der Waals materials. Remarkably high electron distribution in the van der Waals gaps signal strong coupling between layers. We propose that interlayer coupling effect may be primarily responsible for the observed trends in the band gap of 2D InSe. We note that interlayer 
coupling, as a familiar effect in 2D materials\cite{wu_interface_2015,liu_evolution_2014,zhang_observation_2016,chiu_spectroscopic_2014,wang_stacking_2016,fang_strong_2014,zhang_evolution_2015,wang_role_2015,zhang_interlayer_2017,li_probing_2013,chen_observation_2014,dean_hofstadters_2013,yankowitz_emergence_2012,hunt_massive_2013,ponomarenko_cloning_2013} 
has been widely used in detecting layer thickness in multilayer graphene,\cite{wu_interface_2015} 
realizing indirect-to-direct band gap transitions in TMDCs\cite{zhang_evolution_2015,wang_role_2015,zhang_interlayer_2017} 
and Hofstadter's butterfly\cite{dean_hofstadters_2013,yankowitz_emergence_2012,hunt_massive_2013,ponomarenko_cloning_2013} 
in Graphene/$h$-BN heterostructure. Here, we investigate several properties of 2D InSe associated with the interlayer coupling effect, including the magnitudes of band gap, band-edge positions, phonon frequencies, effective masses and carrier mobilities as a function of layer thickness. Our results offer new insight towards understanding the effects of interlayer coupling resulting in tunable band gap and thickness-dependent carrier mobility in few-layer InSe.

\section{\textbf{METHOD}}

Density functional calculations are performed within the generalized gradient approximation (GGA) and Perdew-Burke-Ernzerhof (PBE) exchange-correlation functional\cite{perdew_generalized_1996} as implemented in VASP code.\cite{kresse_efficiency_1996,kresse_efficient_1996,kresse_ab_1993,kresse_ab_1994} 
The electron-ion interaction is described with projector-augmented plane wave (PAW) potentials.\cite{kresse_ultrasoft_1999,blochl_projector_1994} 
For all calculations, the plane-wave energy cutoff is set to 520 eV and the energy convergence threshold is set as $10^{-5} eV$. Monkhorst-Pack k-point meshes with spacing $2\pi \times$ 0.03 $\AA^{-1}$ are employed for the full relaxation of the geometry of 
layered $\beta$-InSe and $\gamma$-InSe and uniform k mesh density is used when calculating band structures. The atomic flake and its neighbouring image are separated by a vacuum space that exceed 15 
$\AA$. For better accuracy, relaxations are performed including the van der Waals (vdW) interaction using optB86b-vdW and 
optB88-vdW functionals.\cite{jiri_klimes_and_david_r_bowler_and_angelos_michaelides_chemical_2010} 
The optimized structures obtained with optB86b-vdW functional for bulk $\beta$-InSe and $\gamma$-InSe 
are both in excellent agreement with the experimental results (ESI Table S1). All subsequent calculations in this study are based on the optB86b-vdW functional. To ascertain the suitability of the DFT with GGA-PBE exchange-correlation, we have performed electronic structure calculations using several different approaches. For samples from monolayer to trilayer $\gamma$-InSe, 
we find that the higher-level hybrid density functional HSE06\cite{krukau_influence_2006} GW+Bethe-Salpeter-Equation (GW+BSE, including excitonic electron-hole interaction)\cite{albrecht_ab_1998,rohlfing_electron-hole_1998}, and GGA-PBE -- all approaches -- present similar trends in electronic band gap variations compared with experimental results (see ESI Figs. S1 and S2). 
Since, our focus is on the variation tendencies of band gaps and band-edge states with layer thickness, we trust that the GGA-PBE results do not affect the conclusions of this study, even though band gaps are underestimated. We calculate the optical phonon modes at the Brillouin Zone center ($\Gamma$) using linear response theory as implemented in the Quantum Espresso package.\cite{paolo_gi_quantum_2009} 
These calculations are performed using norm-conserving pseudopotentials and a kinetic energy cutoff of 100 Ry. Density functional perturbation theory (DFPT)\cite{baroni_phonons_2001} is employed for the study of the vibrational frequencies at $\Gamma$ point for the atomic flakes.

On the basis of effective mass approximation, the carrier mobility in 2D materials is calculated by using the deformation potential (DP) 
theory\cite{bardeen_deformation_1950} and is expressed as:\cite{cai_polarity-reversed_2014,sun_ab_2016,takagi_universality_1994,xu_al2c_2016,wang_reducing_2014,long_theoretical_2009,samantha_bruzzone_ab-initio_2011,xi_first-principles_2012,fiori_multiscale_2013,price_two-dimensional_1981,xie_theoretical_2014}:

\begin{equation}
\mu_{2D} = \frac{e\hbar^3C_{2D}}{k_{B}Tm^*m_d(E_{l}^{i})^2}
\label{equ_1}
\end{equation}

\noindent 

Here, $e$ is the electron charge, $\hbar$ is reduced Planck constant, $k_B$ is Boltzmann constant, and $T$ is the temperature (set to 300 K in this paper). $E_{l}^{i}$ represents the deformation potential constant of the valence band maximum (VBM) for hole or conduction band minimum (CBM) for electron along the transport direction, 
defined as $E_{l}^{i} = \triangle E_{i}/(\triangle l/l_0)$. $\triangle E_{i}$ is energy change of the $i^{th}$ 
band under proper cell dilatation and compression, $l_0$ is lattice constant in the transport direction, and $\triangle l$ 
is the deformation of $l_0$. $m^*$ is effective mass in the transport direction and $m_d$ is the average effective mass 
determined as $m_d = \sqrt{m_xm_y}$. The elastic modulus $C_{2D}$ of longitudinal strain along x and y directions of the 
longitudinal acoustic wave is derived from $(E-E_0)/S_0 = C_{2D}(\triangle l/l_0)^2/2$, where $E$ is the total energy and $S_0$ 
lattice volume at equilibrium for a 2D system.

\section{\textbf{RESULTS and Discussion}}
\subsection{\textbf{Crystalline structure and electronic band structure of few-layer InSe.}}

\begin{figure}[h]
\center
\includegraphics[width=0.48\textwidth]{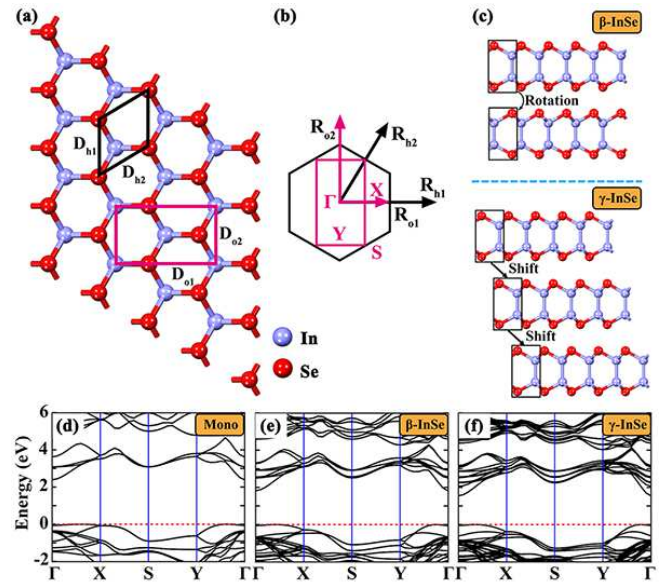}
\caption{(a) Atomic structure of monolayer InSe, where the hexagonal primitive cell (defined by $D_{h1}$ and $D_{h2}$) and orthogonal supercell (defined by $D_{o1}$ and $D_{o2}$) are enclosed with black and magenta frames. (b) The corresponding first Brillouin zones of hexagonal primitive cell and orthogonal supercell. (c) Side views of bilayer $\beta$-InSe (upper panel) and trilayer $\gamma$-InSe (lower panel). The stacking of InSe monolayers represent a rotation and shift behavior for $\beta$-InSe and $\gamma$-InSe, respectively. (d-f) The PBE functional band structures after ``scissors operator'' for monolayer InSe, bilayer $\beta$-InSe, and trilayer $\gamma$-InSe, respectively.}
\label{figure1}
\end{figure}

Fig. \ref{figure1}a shows the top view of monolayer (ML) InSe, a quadruple Se-In-In-Se atomic sheet, where the relevant hexagonal primitive cell and orthogonal supercell are labeled with differently colored (black and magenta) 
frames. The corresponding first Brillouin zones of both primitive and supercell are shown in Fig. 1b. Two distinct structures ($\beta$-InSe and $\gamma$-InSe) with different stacking patterns shown in Fig. 1c are selected for the following explorations. 
$\beta$-InSe and $\gamma$-InSe consist of repeating bilayer and trilayer arrangements in the interlayer direction, respectively. Using first-principles calculations, we demonstrate in the following that the dependence of band gap with varying thickness is mainly dominated by interlayer coupling effect, and its impact on other properties of few-layer InSe.

The calculated band structures of monolayer InSe with GGA-PBE functional and HSE06 functional are shown in ESI Fig. S1. Both functionals achieve consistent results with respect to band ordering and dispersion, but obtain 
different magnitude in the band gaps. The semi-local PBE functional is known to underestimate band gaps. Further test calculations on bilayer 
$\beta$-InSe and trilayer $\gamma$-InSe using GGA-PBE, HSE06, and GW+BSE show similar trends (see ESI Fig. S2). Thus, we believe that PBE functional is able to precisely describe the band characteristics of few-layer InSe, while the band gaps are corrected by applying ``scissors operator'' to rigidly shift the conduction bands (CBs). Similar methodology has been employed in previous studies.\cite{gomes_da_costa_first-principles_1993,parashari_calculated_2008,bernstein_nonorthogonal_2002,johnson_corrections_1998,fiorentini_dielectric_1995,a_thilagam_and_d_j_simpson_and_a_r_gerson_first-principles_2011,ramesh_babu_structural_2011} 
Here the uniform scissors operator, i.e., the band gap difference between the PBE and the HSE06 calculations for monolayer InSe, is adopted. The gap corrected PBE functional band structures of monolayer, bilayer $\beta$-InSe and trilayer $\gamma$-InSe are presented in Fig. 1d-f.

\subsection{\textbf{Interlayer coupling evidenced by the evolution of band gaps and band-edge states with varying thickness of 2D atomic flakes.}}

\begin{figure}[h]
\center
\includegraphics[width=0.48\textwidth]{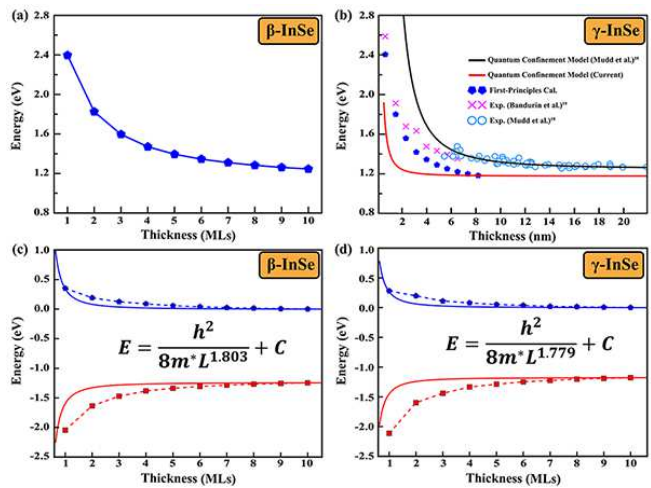}
\caption{(a-b) Evolutions of band gaps as function of layer thickness for $\beta$-InSe and $\gamma$-InSe. In 2b the band gap values from experimental PL data\cite{mudd_tuning_2013,bandurin_high_2017}, the fitting based on current finite potential well model (with varied $\alpha$) and the infinite potential well model (with $\alpha = 2$) from Ref. 18 are shown. We note that the carrier effective mass values (used for calculating exciton reduced mass) in two quantum confinement models are different. (c-d) Evolutions of the band-edge states (CBMs and VBMs) as function of layer thickness for $\beta$-InSe and $\gamma$-InSe. The CBMs and VBMs are plotted with dashed blue and dashed red lines. The fitting formulations of one-dimensional finite potential well models are listed in the figures and the corresponding curves are plotted with solid blue and solid red lines, respectively. The energies of CBMs and VBMs are aligned to the vacuum level and the CBM of 10-MLs $\beta$-InSe and $\gamma$-InSe is set to zero.}
\label{figure2}
\end{figure}

It is known that multilayer graphene flakes have strong in-plane covalent bonds and weak interlayer interactions that show small layer-dependent band gap variations. However, from the $n$-MLs stacked $(In_2Se_2)_n$ (with $n$ = 1-10), 
we find that the calculated gap values remarkably decrease with increasing layer thickness for the two polytypes of InSe as shown in Fig. \ref{figure2}a-b. 
The band gaps are about 2.39 eV for monolayer InSe and decrease to 1.24 eV and 1.18 eV for 10-MLs $\beta$-InSe and $\gamma$-InSe, respectively. The sharp drops of ~1.15 eV and ~1.21 eV is comparable to other strongly coupled interlayer materials.\cite{qiao_high-mobility_2014,zhao_extraordinarily_2016,zhao_high-electron-mobility_2017} 
The high tunability of band gap with varying thickness implies similarly strong interlayer coupling in the two polytypes considered here. Corresponding PBE band structures are listed in ESI Fig. S3 and Fig. S4, respectively. We also compare the calculated results with experimental data extracted from the energy peaks of photoluminescence (PL) spectroscopy\cite{mudd_tuning_2013,bandurin_high_2017} in Fig. 2b. Evidently, our first-principles calculations (after applying scissors correction) agree well with the experimental results.\cite{feng_back_2014,lei_evolution_2014,sanchez-royo_electronic_2014}

To further analyze the band gap diminutions, we assess the evolutions of the heights of the band edges with varying layer thickness as depicted in Fig. \ref{figure2}c-d. The heights of CBMs and VBMs are all aligned with respect to the corresponding vacuum levels, 
while the CBM of 10-MLs InSe is set to 0 eV for easy comparison. From the variation tendencies of CBMs (dashed blue lines) 
and VBMs (dashed red lines), we can see that both band edges contribute to the diminutions of band gaps as layer thickness increases. 
However, the height variations of VBMs are 2.28 and 2.46 times larger than those of CBMs for $\beta$- and $\gamma$-InSe, respectively. This is to say that the band gap variations mainly rely on the upward shift of VBMs.

Previous studies attributed this dependence of the heights of energy levels on layer thickness to quantum confinement effect.\cite{mudd_tuning_2013,yang_spectroscopy_2005,yang_transient_2006} However, we found that a simple classical model based on quantum confinement effect does not appropriately explain the calculated data because the heights of band edges are inconsistent with those predicted with a weak coupling assumption. Under one-dimensional finite potential well model, when the band structures are solely dominated by the quantum confinement effect, the energy levels can be described as: 

\begin{equation}
E = \frac{n^{2}h^{2}}{8m^{*}L^{\alpha}} + C
\label{equ_2}
\end{equation}

\noindent

where, $n$ represents ground ($n = 1$) and excited states ($n > 1$); $h$ is Planck's constant; $m^*$ is the electron or hole effective mass along interlayer direction of bulk $\beta$-InSe and $\gamma$-InSe and $L$ is the well width. Our calculated effective mass values 
($\beta$-InSe: $\arrowvert m_e^* \arrowvert \approx$ 0.043 $m_0$ and $\arrowvert m_h^* \arrowvert \approx$ 0.041 $m_0$, $\gamma$-InSe: 
$\arrowvert m_e^* \arrowvert \approx$ 0.032 $m_0$ and $\arrowvert m_h^* \arrowvert \approx$ 0.035 $m_0$) are similar to those reported in previous works.\cite{gomes_da_costa_first-principles_1993,kuroda_resonance_1980,kress-rogers_cyclotron_1982} In order to estimate $L$, 
we define it as the projection distance along interlayer direction between two sideward Se atoms of InSe flakes with an additional 
1$\AA$ boundary. $\alpha$ is the index (1.803 for $\beta$-InSe and 1.779 for $\gamma$-InSe) and $C$ is a constant for alignment.

We firstly apply the one-dimensional finite potential well model to describe the band gap evolution with the layer thickness for 
$\gamma$-InSe (where the experimental data are available), as shown in Fig. 2b. Here the exciton reduced mass $m_r^*$ = (1/$\arrowvert m_e^* \arrowvert$ + 1/$\arrowvert m_h^* \arrowvert)^{-1}$ is evaluated as the input of $m^*$ in Eq. (\ref{equ_2}). For comparison, another model based on quantum confinement by Mudd et al.,\cite{mudd_tuning_2013} 
is shown in Fig. 2b. Both models fail to describe experimental gaps in the fewer-layer region. The implication is that quantum confinement may not be the sole factor controlling band gap evolution, and thus other effect, i.e., interlayer coupling, has to be included for a complete explanation.

The theoretical heights of CBMs and VBMs based on quantum confinement (i.e., the finite potential well model used here) are plotted as blue and red solid lines in Fig. 2c-d. Although the fitted (solid) curves roughly match the trends of first-principles calculated data (dashed curves), the results gradually deviate in the range of low thickness for both polytypes, indicating again that the one-dimensional finite potential well model is unable to describe the accurate height of band edges. Hence, the interlayer coupling effect should play an important role in the evolution of band gaps, especially the heights of VBMs in both InSe polytypes.

We should point out that although our above results indicate interlayer coupling may be an essential factor in determining the evolution of band-edge states, we should not completely abandon the concept of quantum confinement as the two effects are strongly intertwined, challenging to separate one from the other. This resembles the case of $MoS_2$ where quantum confinement and interlayer coupling coexist and in the fewer-layer region, the interlayer coupling stands out as a strong correction term.\cite{zhang_evolution_2015,mak_atomically_2010,han_band-gap_2011,kang_unified_2016}

\subsection{\textbf{Evolutions of the positions of band-edge states under strong interlayer coupling effect.}}

\begin{figure}[h]
\center
\includegraphics[width=0.48\textwidth]{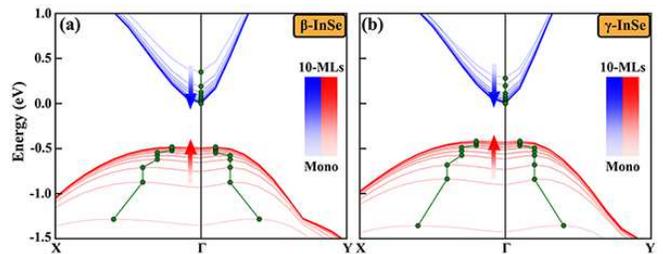}
\caption{Evolution of the lowest CBs (blue lines) and highest VBs (red lines) as function of layer thickness in (a) $\beta$-InSe and (b) $\gamma-$InSe. The color shades become deeper with increasing thickness from monolayer to 10-MLs. The evolution of band-edge states for the lowest CBs at $\Gamma$ and highest VBs along $\Gamma -X$ and $\Gamma -Y$ directions are shown with circular symbols connected by green lines.}
\label{figure3}
\end{figure}

The lowest CB and the highest valence band (VB) of monolayer InSe shown in ESI Fig. S5 depict two nearly degenerate VBMs (within 1 meV) located along $\Gamma -X$ and $\Gamma -Y$ directions, while the CBM is located at $\Gamma$ point. 
This indicates character of indirect band gap transition. From the evolutions of the lowest CBs and the highest VBs from monolayer to 10-MLs flakes as illustrated in Fig. \ref{figure3}, the CBMs remain at $\Gamma$ point and the band-edge shapes demonstrate kinetic energy release effect (strong dispersion) with increasing thickness in both InSe polytypes. The two nearly degenerate VBMs along 
$\Gamma -X$ and $\Gamma -Y$ in monolayer InSe gradually shift towards $\Gamma$ 
point with increasing layer thickness. In other words, interlayer coupling brings a ``lift'' feature at $\Gamma$ point for VBMs that explains the indirect-to-direct band gap transition from monolayer to bulk structures. Similar transition is also observed in Phosphorene that has already been ascribed to strong interlayer coupling.\cite{wang_role_2015}

It should be noted that the effects of interlayer coupling is incorporated in our first principles calculations including weak van der Waals interaction. This is one reason why our calculated band gaps show good agreement with available experimental data (Fig. 2b). Apart from the earlier discussions, the existence of interlayer coupling in 2D InSe is further evidenced by the strong wave-function overlap between the band-edge states of adjacent layers, as shown in ESI Fig. S6. The effect is especially pronounced in the VBM states, consistent with the fact that the substantial deviation between quantum confinement model and first-principles data occurs to the VBM (see Figs. 2c-d).

\subsection{\textbf{Evolutions of the phonon frequencies under strong interlayer coupling effect.}}

Both $\beta$-InSe and $\gamma$-InSe have layer-dependent band gaps with drops of ~1.15 eV and ~1.21 eV from monolayer to 10-MLs, respectively. Such high tunability is the result of orbital hybridization in the interlayer region. In addition to the study of electronic properties, the detections of phonon modes at low frequency regions is already utilized to investigate interlayer interaction in 2D materials. Even in weakly coupled materials such as multilayer graphene, the variations of shear and breathing modes are the ideal tools to gauge flake thickness.\cite{wu_interface_2015}

\begin{figure}[h]
\center
\includegraphics[width=0.48\textwidth]{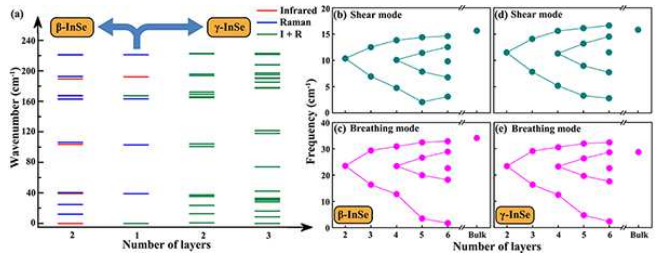}
\caption{(a) Evolutions of phonon frequencies at $\Gamma$ point from monolayer to bilayer $\beta$-InSe and to trilayer $\gamma$-InSe, respectively. The optical activities are expressed with red (Infrared, I), blue (Raman, R) and green (I + R) segments. (b-e) Evolutions of the shear and breathing modes with varying layer thickness for $\beta$-InSe and $\gamma$-InSe, respectively.}
\label{figure4}
\end{figure}

Fig. \ref{figure4}a shows the calculated frequencies and optical activities at $\Gamma$ point for few-layer InSe. Vibration modes are normally classified according to irreducible representation of the crystal symmetry. For the monolayer InSe, it retains a point group symmetry of $D_{3h}$ owing to the presence of the horizontal reflection plane that average the In-In bonds. The corresponding representation is $\Gamma = 2A_2^{''} + 2E^{'} + 2E^{''} + 2A_1^{'}$, where the two $A_2^{''}$ phonon modes are Infrared (I) active, the two 
$E^{'}$ modes are both Raman (R) and Infrared active, and the two $E^{''}$ and two $A_1^{'}$ are both Raman active as shown in ESI Fig. S7. The corresponding symmetry representations and optical activities for bilayer $\beta$-InSe and $\gamma$-InSe  are also illustrated in ESI Fig. S7, where the vibration modes are $\Gamma = 4A_{2u} + 4E_u + 4E_g + 4A_{1g}$ and 
$\Gamma = 12A_1 + 12E$, respectively. $A_{2u}$ and $E_u$ are infrared active, $E_g$ and $A_{1g}$ are Raman active, 
and $A_1$ and $E$ are both Raman and Infrared actives. The differences in optical activities between 
$\beta$-InSe and $\gamma$-InSe could be ascribed to the different stacking patterns that cause the reductions of symmetry operators in $\gamma$-InSe.

Fig. \ref{figure4}b-e shows the calculated shear and breathing modes. The fan-like diagrams signify rapid frequency variations with increasing layer thickness. The separated frequencies retain the unique relationships with layer thickness 
in both $\beta$-InSe and $\gamma$-InSe. In case of shear modes, the calculated frequencies (2-MLs $\to$ 6-MLs) 
gradually descend from 10.37 $cm^{-1}$ to 3.11 $cm^{-1}$ for $\beta$-InSe and from 11.51 $cm^{-1}$ to 2.80 $cm^{-1}$ for 
$\gamma$-InSe, respectively. The breathing mode frequencies vary from 23.51 $cm^{-1}$ to 1.76 $cm^{-1}$ for $\beta$-InSe and from 23.35 $cm^{-1}$ to 2.31 $cm^{-1}$ for $\gamma$-InSe. 
These frequency variations are more than those reported in materials with strong interlayer coupling  ($PtS_2$ and $PtSe_2$),\cite{zhao_extraordinarily_2016,zhao_high-electron-mobility_2017} 
which suggests even stronger effect in InSe flakes. The differences in the breathing and shear mode frequency variations between $\beta$-InSe and $\gamma$-InSe are indicative of distinct interlayer interactions in both polytypes. More specifically, it should be ascribed to the different overlapping of wave functions. As shown in Fig. 4b-e, interlayer coupling induces the fan-like frequency diagrams in both $\beta$-InSe and $\gamma$-InSe, which has already been reported in twisted multilayer graphene with weak coupling.\cite{wu_resonant_2014}

\subsection{\textbf{Evolutions of the carrier mobilities under strong interlayer coupling effect.}}

From an engineering point of view, 2D InSe represents a promising semiconductor for electronic devices, such as the field-effect transistors (FET). Previous experimental studies on few-layer InSe revealed that the electron mobility reaches ~1000 $cm^2V^{-1}s^{-1}$ at room temperature,\cite{feng_back_2014,bandurin_high_2017,segura_electron_1984,feng_back_2014} 
which is the highest reported value among TMDCs\cite{cui_multi-terminal_2015,fallahazad_shubnikov-haas_2016} and is superior to black phosphorus.\cite{li_quantum_2016,qiao_high-mobility_2014} In this regard, the dependence of in-plane mobility on thickness is essential to device performance.

Here we utilize a phonon-limited scattering model to estimate the intrinsic carrier mobilities of InSe at different thicknesses, in which the primary scattering mechanism limiting carrier mobility is lattice phonons\cite{takagi_universality_1994,xu_al2c_2016,wang_reducing_2014,long_theoretical_2009,samantha_bruzzone_ab-initio_2011,xi_first-principles_2012,fiori_multiscale_2013,price_two-dimensional_1981,xie_theoretical_2014} 
The carrier mobilities are not only controlled by the band-edge effective mass, but also by the 2D elastic modulus and the deformation potential. The elastic modulus and the band-edge deformation potential with respect to the vacuum energy are obtained by fitting strain-energy curves, as shown in ESI Fig. S8. All the calculated results of effective mass, deformational potential, 2D elastic modulus and carrier mobility are summarized in ESI Table S2 
($\beta$-InSe) and Table S3 ($\gamma$-InSe)  For InSe, the electrons usually exhibit higher mobility than holes for all thicknesses, because the VBs are generally much flatter than CBs (Fig. 3). The estimated electron mobility for monolayer InSe is 801.09 $cm^2V^{-1}s^{-1}$ along $\Gamma -X$ direction and 689.20 $cm^2V^{-1}s^{-1}$ along $\Gamma -Y$ direction, while the hole mobility is only around 10 $cm^2V^{-1}s^{-1}$.

\begin{figure}[h]
\center
\includegraphics[width=0.48\textwidth]{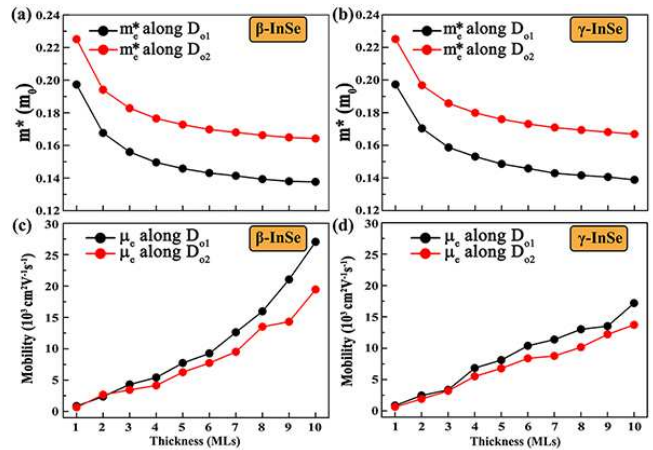}
\caption{Evolutions of (a-b) the electron effective masses and (c-d) the corresponding carrier mobilities along $D_{o1}$ (black lines) and $D_{o2}$ (red lines) directions with increasing layer thickness for $\beta$-InSe and $\gamma$-InSe, respectively.}
\label{figure5}
\end{figure}

Fig. \ref{figure5}a-b shows he trend of inherent electron effective masses ($m^*$) as a function of layer thickness along the $\Gamma -X$ ($D_{o1}$) and $\Gamma -Y$ ($D_{o2}$) directions. 
vidently, they gradually become small with increasing thickness, which is associated with the release process at $\Gamma$ point as the well-width increases. The electron conduction exhibits slightly anisotropic behavior, since the effective masses along $\Gamma -X$ 
are generally smaller than those along $\Gamma -Y$ direction at all thickness for both $\beta$- and $\gamma$-InSe. 
Fig. 5c-d shows the corresponding results of electron mobilities. As expected, they consistently increase with thickness. It is notable that the mobilities of 
$\beta$-InSe are always higher than those of $\gamma$-InSe at any given thickness, which may be more favorable for electronic applications. As compared with the value of monolayer InSe, the mobility reaches about 10-30 k 
$cm^2V^{-1}s^{-1}$ for 10-MLs, exhibiting an increase of 1-2 orders of magnitude. Such a large increase in mobility results mainly from the hardening of lattice caused by remarkably higher 2D elastic modulus of multi-layer InSe (see ESI Table S2 and Table S3). In this respect, we deduce that interlayer coupling is mainly responsible for the variation of strength of phonon scattering and would be a potential parameter available for mobility engineering in multi-layer 2D semiconductors.

\section{\textbf{CONCLUSIONS}}

By employing ab initio calculations including long-range dispersion interaction, we analyzed the evolution of electronic energies in $n$-MLs stacked $(In_2Se_2)_n$ with number of layers $n $ = 1-10. 
The layer-dependent band gap values exhibit sharp drops ~1.15 eV and ~1.21 eV from monolayer to 10-MLs for $\beta$-InSe and $\gamma$-InSe, respectively. 
A fitting with one-dimensional finite potential well model does not appropriately match the evolution of VBMs obtained by first-principles methods, implying that the changes in the band gap with layer thickness may not be dominated by quantum confinement effect. We surmise that strong interlayer coupling could be the primary cause which is manifested as following: first, the emergence of indirect-to-direct band gap transitions with increasing layer thickness; second, the fan-like frequency diagrams of shear and breathing modes and their variations with layer thickness derived from different interlayer interactions; finally, the layer-dependent carrier mobilities, whose values increase from about 0.80 k 
$cm^2V^{-1}s^{-1}$ for the shared monolayer structure to 10-30 k $cm^2V^{-1}s^{-1}$ for 10-MLs $\beta$-InSe and $\gamma$-InSe, respectively. The considerably higher carrier mobilities of $\beta$-InSe and $\gamma$-InSe along with their reduced band gaps should be desirable in electronic applications. With high band gap tunability and excellent transport properties, we speculate that $\beta$-InSe and $\gamma$-InSe as strongly coupled layered 2D structures possess significant potential and deserve attention from FET industry.

\begin{acknowledgments}
The authors acknowledge funding support from the National Natural Science Foundation of China (under Grant Nos. 11404131, 11674121, 61674080, and 61722403), Program for JLU Science and Technology Innovative Research Team, and the Special Fund for Talent Exploitation in Jilin Province of China. KB acknowledges support from US Department of Homeland Security under Grant Award Number, 2014-DN-077-ARI075.
\end{acknowledgments}

\bibliographystyle{rsc}
\bibliography{reference}

\end{document}